\documentclass[twocolumn]{jpsj3} 

\usepackage{times}

\usepackage{mathrsfs}

\newcommand{\diff}{\mathrm{d}}
\newcommand{\imag}{\mathrm{Im}\,}

\newcommand{\imu}{\mathrm{i}}

\usepackage{amsmath,amsbsy,amssymb,graphics}
\usepackage{bm}

\title{
Resolution of Entropy $\ln\sqrt 2$ by 
Ordering in Two-Channel Kondo Lattice
}

\author{
Shintaro \textsc{Hoshino}$^{1}$\thanks{E-mail: hoshino@cmpt.phys.tohoku.ac.jp},
Junya \textsc{Otsuki}$^{1,2}$ and 
Yoshio \textsc{Kuramoto}$^{1}$
}

\inst{
$^{1}$Department of Physics, Tohoku University, Sendai 980-8578, Japan \\
$^{2}$Theoretical Physics III, Center for Electronic Correlations and Magnetism, Institute of Physics, \\
University of Augsburg, Augsburg 86135, Germany
}
\recdate{\today}

\abst{
Peculiar property of electronic order is clarified for the two-channel Kondo lattice.  With two conduction electrons per site, the order parameter is a composite quantity involving both local and itinerant degrees of freedom.
In contrast to the ordinary Kondo lattice, a heavy electron band is absent above the transition temperature, but is rapidly formed below it.  
The change of entropy associated with the ordering is found to be close to $\ln \sqrt 2$ per site.  This entropy corresponds to the residual entropy in a two-channel Kondo impurity, which has been regarded as due to localized free Majorana particles.  
The present composite order is interpreted as instability of 
Majorana particles toward non-Kramers conduction electrons plus 
heavy fermions that involve localized electrons. 
}

\kword{
two-channel Kondo lattice,
composite order,
Majorana fermion,
dynamical mean-field theory,
continuous-time quantum Monte Carlo
}

\begin{document}
\maketitle

\section{Introduction}

Non-Kramers systems with even number of $f$ electrons per site as in Pr$^{3+}$ and U$^{4+}$ are expected to show intriguing phenomena that is different from ordinary Kramers systems with odd number of electrons in Ce- and Yb-based systems.
Under the cubic crystalline electric field, the localized $f^2$-electron state can have a non-Kramers doublet ground state where $f$ electrons do not have spin degrees of freedom.
This doublet interacts through orbital with conduction electrons which have spin and orbital degrees of freedom.
Thus the two-channel Kondo system can be realized \cite{cox98}.
Experimentally, several candidates such as U$_{x}$Th$_{1-x}$Ru$_{2}$Si$_{2}$ and Pr$_{x}$La$_{1-x}$Ag$_{2}$In have been reported \cite{amitsuka94, kawae05}.
Recently the two-channel Kondo systems have attracted more attention since the discovery of a series of non-Kramers doublet compounds such as PrIr$_2$Zn$_{20}$ \cite{onimaru10,onimaru11} and PrTi$_2$Al$_{20}$ \cite{sakai11}.

\begin{figure}[b]
\begin{center}
\includegraphics[width=70mm]{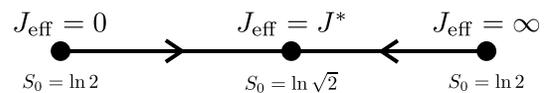}
\caption{
Renormalization flow of the impurity two-channel Kondo model.
}
\label{fig_scaling}
\end{center}
\end{figure}

In the two-channel Kondo impurity, where the localized spin is located on a metal with degenerate channel degrees of freedom,  the scaling theory tells us that the weak and strong coupling limits does not become the fixed point \cite{cox98}.
The system instead chooses the intermediate effective coupling $J_{\rm eff} = J^*$ as the fixed point, which is in strong contrast with the ordinary Kondo model.
This scaling flow is illustrated in Fig.~\ref{fig_scaling}.
The ground state has the peculiar residual entropy $S_0=\ln\sqrt 2 $ corresponding to a part of localized spin.
This remaining degrees of freedom is interpreted as the localized Majorana fermions \cite{emery92}.
In the periodically aligned $f$-electron systems, on the other hand, this partially removed but remaining entropy will lead the system to highly non-trivial phase transitions by inter-site interactions.

The simplest description of the periodic non-Kramers doublets coupled to conduction electrons is given by the two-channel Kondo lattice (KL) \cite{jarrell96}.
One of the characteristics in this model is the channel symmetry breaking, which is absent in the ordinary KL.
At one conduction electron per site (quarter filling), the possibility of the channel symmetry breaking with staggered alignment is pointed out \cite{cox98, schauerte05}.
The stabilization of this phase at finite temperatures is demonstrated in infinite dimensions \cite{hoshino_iche}.
At two conduction electrons per site (half filling), on the other hand, the dynamical mean-field study has pointed out the channel symmetry breaking using exact diagonalization with finite sized bath \cite{nourafkan08}.

Recently we have demonstrated that the order parameter for the channel-symmetry-broken phase at half filling is the composite quantity involving both localized spin and conduction electrons \cite{hoshino11}.
In this paper, we investigate the details of this ordered state.
By making a comparison with conventional ordered state, we demonstrate that the entropy resolved below the transition temperature is related to $S_0=\ln\sqrt 2$ at the non-trivial fixed point shown in Fig.~\ref{fig_scaling}.
We argue that this entropy also implies the relevance of the Majorana fermions in the two-channel KL.

The two-channel KL reads
\begin{align}
{\cal H} &= \sum _{\bm{k} \alpha \sigma} ( \varepsilon _{\bm{k}} -\mu ) c_{\bm{k}\alpha \sigma}^\dagger c_{\bm{k}\alpha \sigma}
+ J \sum_{i} \bm{S}_i \cdot ( \bm{s}_{{\rm c}i1} + \bm{s}_{{\rm c}i2} )
, \label{eqn_2ch_KLM}
\end{align}
where $c_{\bm{k}\alpha \sigma} (c_{\bm{k}\alpha \sigma}^\dagger)$ is annihilation (creation) operator of the conduction electron with channel $\alpha =1,2$ and pseudo-spin $\sigma = \uparrow, \downarrow$.
The operators $\bm{S}_i$ and $\bm{s}_{{\rm c}i\alpha}=\frac{1}{2}\sum_{\sigma\sigma'} c^\dagger_{i\alpha\sigma} \bm{\sigma}_{\sigma\sigma'}c_{i\alpha\sigma'}$ are the localized and conduction pseudo-spins at site $i$, respectively.
Here $\bm{\sigma}$ is the spin-$1/2$ Pauli matrix.
The Hamiltonian (\ref{eqn_2ch_KLM}) has the $SU(2)$ symmetry in both the 
pseudo-spin and channel spaces.
The two-channel KL is applied to the non-Kramers doublet system; 
we regard pseudo-spins as orbital degrees of freedom, and channels as Kramers doublet originating from time-reversal symmetry for conduction electrons.  Hence we write the symmetry as
$SU(2)_O$ for pseudo-spin and $SU(2)_T$ for channels.
In the following, we refer to pseudo-spin simply as `spin' except in \S\ref{subsec_non-Kramers}.

For finite-temperature analysis, we have used the dynamical mean-field theory (DMFT) \cite{georges96} which becomes the exact theory in infinite dimensions.
The semi-circular density of states defined by
$\rho_0 (\varepsilon) = (2/\pi) \sqrt{1 - (\varepsilon / D) ^2}$
is taken for conduction electrons.
We assume the nesting property $\varepsilon_{\bm{k}+\bm{Q}} = - \varepsilon_{\bm{k}}$ with $\bm Q$ being one of the nesting vectors.
For numerical method as impurity solver, we employ the continuous-time quantum Monte Carlo method (CTQMC) \cite{gull11}.
The algorithm we have used is similar to the one explained in ref. \citen{hoshino10}.
In the literature, the CTQMC is developed for the model where two localized spins interact with two conduction bands.
When these two localized spins are regarded as identical, the model is the same as the two-channel Kondo model.
Analytic continuation from imaginary Matsubara frequencies onto real ones is performed by the Pad\'{e} approximation.

The present paper is organized as follows.
We investigate in \S\ref{sec_para_state} the properties of the disordered states.
The phase diagram of the two-channel KL is shown in \S\ref{sec_phase}.
Sections \S\ref{sec_f-channel} and \ref{sec_compari} are devoted to discussions about the symmetry broken phases.
We discuss possible applications in \S\ref{sec_discussion}, and summarize the results in \S\ref{sec_summary}.

\section{Thermodynamics in Paramagnetic state}\label{sec_para_state}

Before we consider the symmetry-broken phases, we study basic thermodynamic properties
of the paramagnetic state.
From the transport analysis by the DMFT, Jarrell {\it et al} have shown that the paramagnetic state is an incoherent metal \cite{jarrell96}.
In this section, we investigate this phase in terms of thermodynamic quantities such as the susceptibility and specific heat.

\subsection{Susceptibilities}

In general, the susceptibility of the operator ${\cal M}$ is given by
\begin{align}
\chi = \int^{1/T}_{0} \left[ \langle {\cal M} (\tau) {\cal M}^\dagger \rangle 
-\langle {\cal M} \rangle ^2
\right] \diff \tau
,
\end{align}
where ${\cal M} (\tau) = e^{\tau {\cal H}} {\cal M} e^{-\tau {\cal H}}$ is the Heisenberg picture with imaginary time.
As the choice of ${\cal M}$, we consider the charge, spin, channel and spin-channel moments of conduction electrons:
\begin{align}
\hat m_{\rm charge} (\bm{q}) &= N^{-1} \sum_{i\alpha\sigma} 
n_{i\alpha\sigma} \,  e^{-\imu \bm{q} \cdot \bm{R}_i}
, \label{eq_mom_charge}\\
\hat m_{\rm spin} (\bm{q}) &= N^{-1} \sum_{i\alpha\sigma} 
\sigma^z_{\sigma\sigma} n_{i\alpha\sigma} \, e^{-\imu \bm{q} \cdot \bm{R}_i}
, \label{eq_mom_spin}\\
\hat m_{\rm chan} (\bm{q}) &= N^{-1} \sum_{i\alpha\sigma} 
\sigma^z_{\alpha\alpha} n_{i\alpha\sigma}  \, e^{-\imu \bm{q} \cdot \bm{R}_i}
, \label{eq_mom_chan}\\
\hat m_{\rm spin\mathchar`-chan} (\bm{q}) &= N^{-1} \sum_{i\alpha\sigma} 
\sigma^z_{\alpha\alpha} \sigma^z_{\sigma\sigma}  n_{i\alpha\sigma} \,  e^{-\imu \bm{q} \cdot \bm{R}_i}
, \label{eq_mom_spin-chan}
\end{align}
where the number of sites is given by $N$.
We have introduced the local number operator $n_{i\alpha\sigma} = c_{i\alpha\sigma}^\dagger c_{i\alpha\sigma}$.
We also consider the localized-spin moment defined by 
\begin{align}
\hat M_{\rm spin} (\bm{q}) = N^{-1} \sum_{i} S_i^z e^{-\imu \bm{q} \cdot \bm{R}_i}
, \label{eq_mom_Spin}
\end{align}
where $S_i^z$ is the $z$ component of the localized spin.
We note that $f$ electrons have only the spin susceptibility.

\begin{figure}
\begin{center}
\includegraphics[width=70mm]{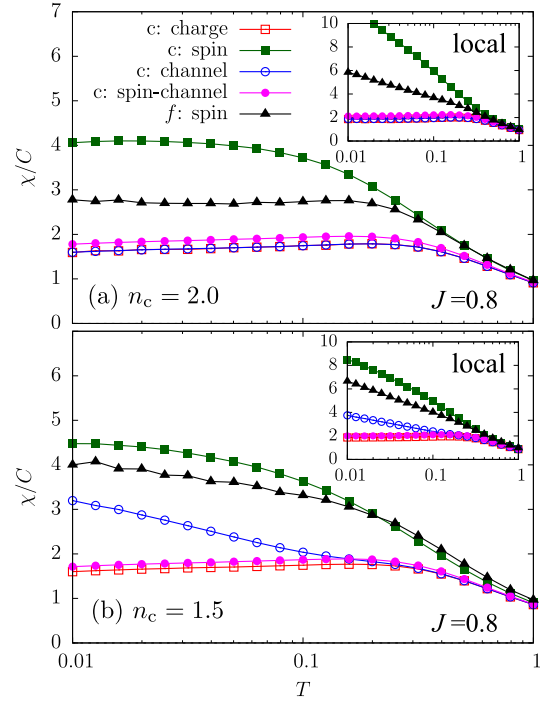}
\end{center}
\caption{(Color online)
Uniform spin and channel susceptibilities in the paramagnetic phase for (a) $n_{\rm c} = 2.0$ and (b) $n_{\rm c} = 1.5$.
The insets show the local susceptibilities.
}
\label{fig_2ch_thermodynamics}
\end{figure}

Figure \ref{fig_2ch_thermodynamics} shows the temperature dependence of the uniform susceptibilities $\chi$ with $\bm{q} = \bm{0}$ for two cases of
the number of conduction electrons $n_{\rm c}=2$ and 1.5 per site.
Each $\chi$ is normalized by the respective Curie-like constant $C$ so that 
$\chi \rightarrow 1/T$ at high temperatures.
At $n_{\rm c} = 2$ shown in Fig.~\ref{fig_2ch_thermodynamics}(a), 
the uniform susceptibilities keep constant values even at low temperatures.
In contrast, local susceptibilities displayed in the inset show the logarithmic temperature dependence.
On the other hand, at $n_{\rm c} = 1.5$ in Fig.~\ref{fig_2ch_thermodynamics}(b), 
the uniform susceptibilities also show the $\ln T$ behavior.
It seems that the particle-hole (p-h) symmetry suppresses
the anomalous $\ln T$ term in the uniform susceptibility.
The apparent Fermi-liquid like behavior is also observed in the specific heat as discussed next.

\subsection{Specific heat} \label{subsec_spec}

\begin{figure}
\begin{center}
\includegraphics[width=70mm]{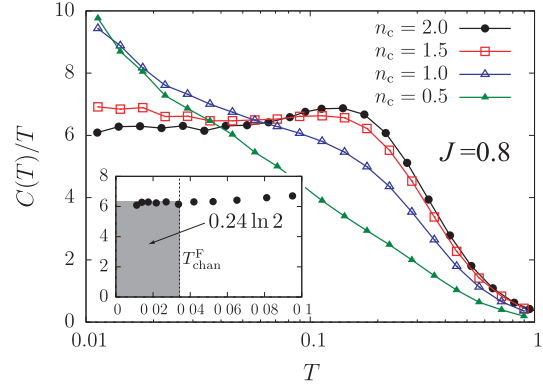}
\end{center}
\caption{
Temperature dependence of the specific heat in the paramagnetic state at various fillings.
The inset is the plot at $n_{\rm c} = 2$ with linear scale for $T$.
More details are discussed in \S\ref{subsec_internal_ene}.
}
\label{fig_2ch_spec}
\end{figure}

The specific heat $C$ in the paramagnetic state is shown in Fig.~\ref{fig_2ch_spec}, which is calculated by differentiating the internal energy with respect to temperature.
The details of numerical derivation of the internal energy are given in Appendix.
At $n_{\rm c} = 2$, $C/T$ shows a slight peak at $T\simeq 0.15$ and tends to a constant for $T \lesssim 0.06$ within the limit of accuracy.
Note that
the fixed-point model (Toulouse limit) of the two-channel Kondo impurity shows the same behavior
\begin{align}
C = \gamma T
,
\end{align}
where $\gamma =\pi/(6\Gamma)$ with $\Gamma$ corresponding to the resonance width of the Majorana fermion\cite{emery92}.  The value of $\gamma$ is half of an ordinary fermion system with the same $\Gamma$.
Away from half filling, on the other hand, the specific heat shows the logarithmic behavior as in the case of uniform susceptibilities shown in Fig.~\ref{fig_2ch_thermodynamics}(b).

Let us discuss the entropy that remains in the hypothetical paramagnetic ground state at half filling.
By comparing the entropy of an ordered state that has zero entropy at the ground state, we can derive the residual entropy.
Relegating the reasoning to \S\ref{subsec_internal_ene},
we quote the final result for the residual entropy
\begin{align}
S(T=0) \simeq 0.55\ln 2,
\label{residual}
\end{align}
per site.
This value is very close to $S_0=\ln\sqrt 2$ that corresponds to the impurity system.

\section{Phase Diagram and 
Strong-Coupling Limit
}\label{sec_phase}

\begin{figure}
\begin{center}
\includegraphics[width=85mm]{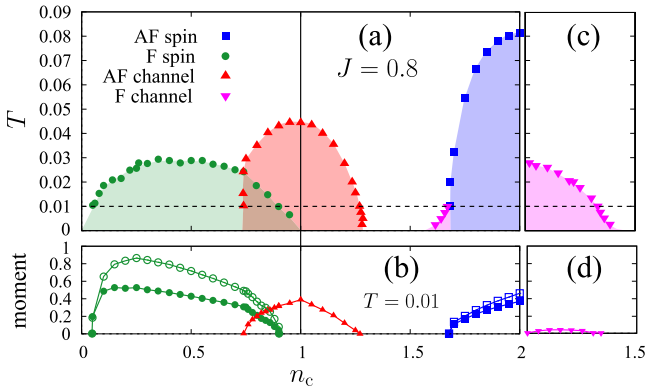}
\caption{
(Color online)
(a) Phase diagram of the two-channel KL with $J=0.8$ and (b) ordered moments at $T=0.01$.
In (b), the filled symbols show the moment of localized spin, and the blank symbols for conduction electrons.
(c) Phase diagram near half filling without staggered orders, and (d) channel moment at $T=0.01$.
The full moment in normalized to unity in (b) and (d).
}
\label{fig_phase}
\end{center}
\end{figure}
\begin{figure}
\begin{center}
\includegraphics[width=85mm]{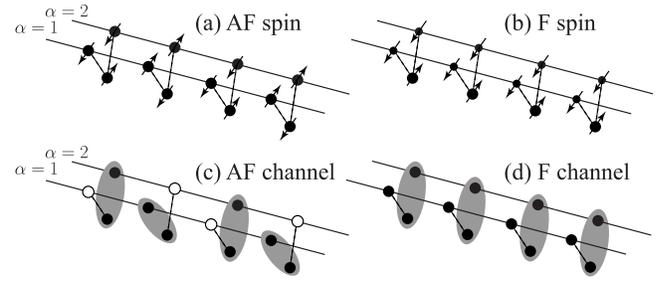}
\caption{
Schematic illustrations for the ordered states in the two-channel KL.
Electrons on line show conduction electrons, and the shaded ovals show the formation of the Kondo singlet.
The filled and open symbols show the presence and absence of electrons, respectively.
The size of the filled circle on the line shows the occupation number of conduction electrons.
}
\label{fig_illust}
\end{center}
\end{figure}

Now we discuss the ordered states.
The phase diagram of the two-channel KL for $J=0.8$ is shown in Fig.~\ref{fig_phase}(a), which is derived from the divergence of the susceptibilities \cite{georges96}.
Here we denote the ordering vectors $\bm{q} = \bm{0}$ and $\bm{Q}$ by F (ferro) and AF (antiferro), respectively.
When we take the other values of $J$, there are only quantitative changes.
As expected from the residual entropy $\ln\sqrt 2$ in the impurity system, the two-channel KL shows ordering in a wide parameter range.
Figure \ref{fig_phase}(b) shows the order parameters of each phase at $T=0.01$, which are defined in eqs.~(\ref{eq_mom_charge})--(\ref{eq_mom_Spin}).
The magnitudes of moments behave in a manner similar to the transition temperature.
The schematic pictures of these ordered states are illustrated in Fig.~\ref{fig_illust}.

The phase around $n_{\rm c} = 2$ is the antiferro(AF)-spin order that has been discussed in ref. \citen{jarrell97}.
We see also the ferro(F)-spin ordered phase with $\bm{q} = \bm{0}$ in the dilute conduction electron region.
These spin orders are found also in the ordinary KL, and are not specific to the two-channel case.
On the other hand, we have found the AF-channel order around $n_{\rm c} = 1$ and F-channel order around $n_{\rm c} \sim 1.6$.
The channel-symmetry breaking is a characteristic of the two-channel model, and cannot be explained by the RKKY interaction since the Hamiltonian (\ref{eqn_2ch_KLM}) includes only the interaction between spins.
Especially the F-channel order will be discussed in greater detail in this paper because of its peculiar and interesting properties.

The AF orders in the phase diagram of Fig.~\ref{fig_phase}(a) can be qualitatively understood from the atomic limit whose Hamiltonian is given by
\begin{align}
{\cal H}_{\rm atomic} =
-\mu \sum _{\alpha \sigma} n_{\alpha \sigma}
+ J \bm{S} \cdot (\bm{s}_{{\rm c}1} + \bm{s}_{{\rm c}2}),
\end{align}
where we omit the site index.
The number $n_{\rm c}$ of conduction electrons is adjusted by the chemical potential $\mu$.
At $n_{\rm c}=2$, the ground state is the doubly degenerate states with the total spin $1/2$.   
The degeneracy is lifted by the inter-site interaction to cause the AF-spin 
order.
On the other hand, the ground state at $n_{\rm c}=1$ is the singlet state formed between localized and conduction spins.
However, conduction spins also have
channel degrees of freedom, and the ground state is again doubly degenerate.
The degeneracy is lifted by the inter-site interaction to cause the 
AF-channel order \cite{cox98}.
At $n_{\rm c}=0$, we have the free localized spin at each site.
With finite but small concentration of conduction electrons, they mediates the interaction between localized spins to form the F-spin ordering as in the ordinary KL.

We comment on the region around $n_{\rm c}\sim 1.5$ in Fig.~\ref{fig_phase}(a), where the disordered state seems to persist down to $T=0$.
To remove the finite entropy, the system should show some ordering at sufficiently low temperature, but its transition temperature can be tiny.
Note that the local spin fluctuation remains as is clear from the logarithmic temperature dependence of the susceptibility in Fig.~\ref{fig_2ch_thermodynamics}.

\section{Nature of Composite Order }\label{sec_f-channel}

\subsection{Growth of order parameters} \label{sec_composite}

Interestingly, the picture from the strong coupling limit does not work
for the F-channel order.
In order to investigate this ordering in detail, we assume that AF orders are strongly suppressed.
Such a situation is expected in the realistic system because of {\it e.g.} next-nearest-neighbor hopping or geometrical frustration.
As shown in Fig.~\ref{fig_phase}(c), the transition temperature $T_{\rm chan}^{\rm F}$ of the F-channel phase increases with approaching to half filling.
The channel moment at $T=0.01$ shown in Fig.~\ref{fig_phase}(d) is finite but tiny. 
At the maximum transition temperature of the F-channel order at $n_{\rm c} = 2$, the channel moment vanishes.
Hence, the channel moment is not a proper order parameter.

We have identified \cite{hoshino11} that a 
composite quantity defined by
\begin{align}
\hat \Psi_i = \bm{S}_i \cdot ( \bm{s}_{{\rm c}i1} - \bm{s}_{{\rm c}i2} ), 
\label{eq_comp_def}
\end{align}
for each site $i$ serves as the proper order parameter.
In the Fourier space, the order parameter is given by $\langle \hat \Psi_{\bm{q}=\bm{0}} \rangle$ where
\begin{align}
{\hat \Psi}_{\bm{q}} = N^{-1} \sum_i \hat\Psi_i e^{-\imu \bm{q}\cdot \bm{R}_i}
.
\end{align}
We note that the composite order is not described by the one-body mean fields such as $\langle \bm{S}_i \rangle$ and $\langle \bm{s}_{{\rm c}i1} - \bm{s}_{{\rm c}i2} \rangle$.
The expression (\ref{eq_comp_def}) is known as the channel-symmetry-breaking perturbation in the two-channel Kondo impurity \cite{affleck92, mitchell12, mitchell12-2}.
In the present case, this term appears by spontaneous symmetry breaking.

Let us first consider the system at half filling.
In addition to $\langle \hat \Psi_{\bm{q}=\bm{0}} \rangle$ 
the kinetic energy and double occupation become dependent on channel below the transition temperature, which is first pointed out in ref. \citen{nourafkan08}.
In order to characterize these quantities, we introduce the normalized order parameter by
\begin{align}
{\bar \phi} \equiv \left| \frac{\phi_1 - \phi_2}{\phi_1 + \phi_2} \right| \leq 1
,\label{eq_norm_para_def}
\end{align}
where 
the channel-dependent quantity $\phi_\alpha \ (\alpha =1,2)$ is chosen as:
\begin{align}
\phi_\alpha = 
\left\{
\begin{matrix}
{\rm (i)} & \langle \bm{S}_i \cdot \bm{s}_{{\rm c}i\alpha}\rangle
\\[1.5mm]
{\rm (ii)} &N^{-1}\sum_{\bm{k}\sigma} \varepsilon_{\bm{k}} \langle c_{\bm{k}\alpha \sigma}^\dagger c_{\bm{k}\alpha \sigma}\rangle
\\[1.5mm]
{\rm (iii)} & \langle (n_{i\alpha\uparrow}-\frac{1}{2})(n_{i\alpha\downarrow}-\frac{1}{2}) \rangle 
\\[1.5mm]
{\rm (iv)} & \sum_{\sigma} \langle n_{i\alpha\sigma}\rangle 
\end{matrix}
\right.
\label{eq_order_prms}
\end{align}
For the double occupation (iii), one could choose $n_{i\alpha\uparrow}n_{i\alpha\downarrow}$ instead of eq.~(\ref{eq_order_prms}).
However, this expression does not become the order parameter under the channel field as will be discussed in \S\ref{subsec_sym_ana}.
Hence we need to subtract $1/2$ from the local number operator as in (iii) of eq.~(\ref{eq_order_prms}) for natural definition of order parameter.

\begin{figure}
\begin{center}
\includegraphics[width=75mm]{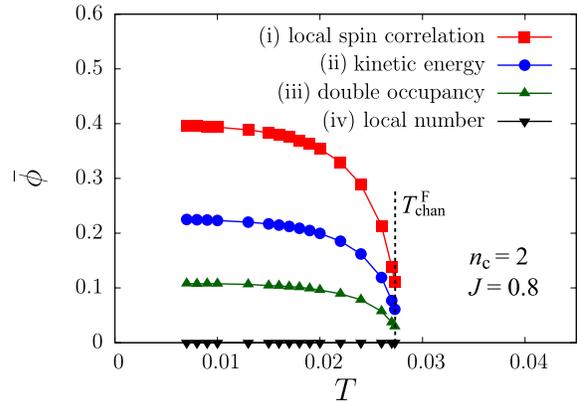}
\caption{
(Color online)
Normalized order parameter $\bar \phi$ defined in eqs. (\ref{eq_norm_para_def}) and (\ref{eq_order_prms}) as a function of temperature.
The full moment is unity.
}
\label{fig_normal_op}
\end{center}
\end{figure}

Figure \ref{fig_normal_op} shows the growth of the order parameter $\bar \phi$ as a function of temperature.
The composite order parameter (i) is the largest among the order parameters, and the kinetic energy (ii) and double occupation (iii) have the smaller values.
Since the coupling between localized and conduction spins is a driving force of the order, the primary order parameter should be $\langle \hat \Psi_{\bm{q}=\bm{0}} \rangle$ and the resultant imbalance of channels is seen in kinetic energies and double occupations as a secondary effect.
Indeed, as seen in \S\ref{subsec_internal_ene}, the kinetic energy increases by the phase transition, which means that the F-channel order is disadvantageous for conduction electrons.
The interaction energy is lowered by the phase transition.
When we focus only on the conduction electrons, the quantity (ii) is relevant as the order parameter.
It is interpreted as the channel ordering in $\bm{k}$ space.

The order parameter $\langle \hat \Psi_{\bm{q}=\bm{0}} \rangle$ is a composite quantity, and cannot be described by any one-body mean field.
Instead the effective mean-field Hamiltonian of the F-channel phase is given by
\begin{align}
{\cal H}_{\rm MF} = {\cal H} + \Delta \sum_{i} \hat \Psi_{i}
. \label{eq_eff_mean_hamilt}
\end{align}
Here $\Delta$ is the magnitude of the {\it two-body mean field}.
Thus, even in the mean-field picture, we still have to solve the many-body problem for the F-channel ordering, which is in strong contrast to conventional orders.

\subsection{Symmetry analysis } \label{subsec_sym_ana}

Respecting the $SU (2)_{T}$ symmetry in the channel space,
the composite order parameter can also be written in the
symmetric form as \cite{affleck92}
\begin{align}
\hat {\bm{\Psi}}_i = \frac{1}{2} \sum_{\alpha\alpha'}\sum_{\sigma\sigma'}
c_{i\alpha\sigma}^\dagger (\bm{S}_i \cdot \bm{\sigma}_{\sigma\sigma'} ) \bm{\sigma}_{\alpha\alpha'}c_{i\alpha'\sigma'}
\equiv 
\bm{S}_i \cdot \hat{\bm t}_i,
\end{align}
where we have introduced the spin-channel moment tensor by
\begin{align}
\left( 
\hat{\bm t}_i \right)^{\mu\beta} = \frac 12
\sum_{\alpha\alpha'}\sum_{\sigma\sigma'}
c_{i\alpha\sigma}^\dagger {\sigma}^\mu_{\sigma\sigma'} 
\sigma_{\alpha\alpha'}^\beta c_{i\alpha'\sigma'},
\label{spin-channel}
\end{align}
which can be regarded as a vector either in spin or channel space. 
Especially $( \hat{\bm t}_i )^{zz}$ has the same form as eq.~(\ref{eq_mom_spin-chan}).
The $z$ component of the vector $\hat {\bm{\Psi}}_i$
coincides with the operator defined in eq.~(\ref{eq_comp_def}).
It is clear that the order parameter breaks the $SU(2)_T$ (time-reversal) symmetry.  
Since $SU(2)_O$ symmetry is {\it not} broken, the orbital Kondo effect can be active in the ordered phase, and can make the pseudo-spin (orbital) singlet.

In terms of symmetry, 
the channel moment $\hat {\bm m}_{\rm chan}$, which includes 
$\hat m_{\rm chan}$ as the $z$-component 
is also a vector in the $SU(2)_{T}$ space.  Hence
$\hat \Psi$ and $\hat m_{\rm chan}$ are in general mixed.
If the p-h symmetry is present, however, they are decoupled.
Let us discuss the situation using the Landau free energy,
which describes the F-channel order as
\begin{align}
F = A \Psi^2 + B\Psi^4 + I\,  m_{\rm chan} \Psi + a \, m_{\rm chan}^2
, \label{eq_free_energy_ph}
\end{align}
where 
$I, B(>0)$ and $a(>0)$ are constants, while $A$ is a linear function of temperature.
The transition temperature is given by the condition $A=I^2/4a$.

We shall show that $I=0$ if the p-h symmetry is present. 
The p-h transformation $\mathscr{P}$, involving the localized states, is defined by
\begin{align}
\mathscr{P} c_{i\alpha \sigma} \mathscr{P}^{-1} &= 
c_{i\alpha \sigma}^\dagger
, \label{eq_ph_trans_1}  \\
\mathscr{P} c^\dagger_{i\alpha \sigma} \mathscr{P}^{-1} &= 
c_{i\alpha \sigma}
, \label{eq_ph_trans_2}  \\
\mathscr{P} |i\sigma \rangle \langle i \sigma'| \mathscr{P}^{-1} &= \delta_{\sigma\sigma'} - |i\sigma'\rangle \langle i\sigma|
, \label{eq_ph_trans_x}
\end{align}
where $|i\sigma\rangle$ is a localized-spin state at site $i$.
Hence $\Psi$ as well as 
the interaction part given by eq.~(\ref{eqn_2ch_KLM}) is
invariant under $\mathscr{P}$.  However 
we obtain the sign reversal for one-body quantities such as the kinetic energy and the channel moment:
\begin{align}
\mathscr{P} \hat m_{\rm chan} (\bm{q}\hspace{-1mm}=\hspace{-1mm}\bm{0}) \mathscr{P}^{-1} &= - \hat m_{\rm chan} (\bm{q}\hspace{-1mm}=\hspace{-1mm}\bm{0})
\end{align}
The conduction electrons are p-h symmetric at $n_{\rm c}=2$, since the chemical potential is at the center of the band.
Hence, after operating $\mathscr{P}$, the free energy $F$ remains the same, while $m_{\rm chan}$ changes sign. 
Then $I$ should be zero in this case.  Namely,
the channel moment may remain zero in the p-h symmetric case.
Away from half filling in the F-channel order, the tiny channel moment arises 
as seen in Fig.~\ref{fig_phase}(d).

The condition $I =0$ with the p-h symmetry suggests that 
the spontaneously broken symmetry in the F-channel phase may also
involve the p-h symmetry.
In order to identify the relevant symmetry of the 
composite order,
we introduce an external field defined by
\begin{align}
{\cal H}_{\rm ext} = - \frac{1}{2} hN
\hat m_{\rm chan} (\bm{q}\hspace{-1mm}=\hspace{-1mm}\bm{0}) 
, \label{eq_channel_field}
\end{align}
which breaks the channel symmetry, and hence $SU(2)_T$.
\begin{figure}
\begin{center}
\includegraphics[width=70mm]{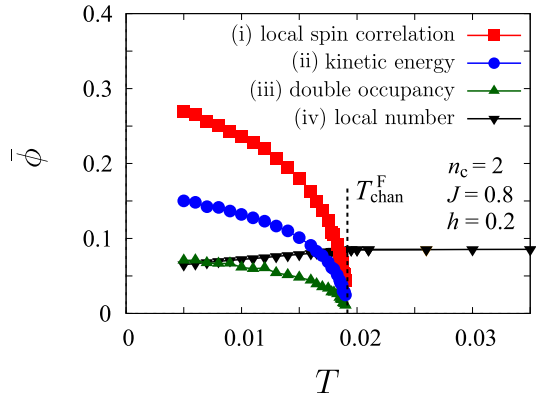}
\caption{
(Color online)
Order parameters and channel moment at $h=0.2$ as a function of temperature.
The definitions of the symbols are the same as Fig.~\ref{fig_normal_op}.
}
\label{fig_channel_field}
\end{center}
\end{figure}
Figure \ref{fig_channel_field} shows the calculated order parameter defined in eqs.~(\ref{eq_norm_para_def}) and (\ref{eq_order_prms}) for $h=0.2$.
The channel moment is finite even at high temperatures because of the 
external field.
However, the composite order parameter $\Psi$ becomes finite only below $T_{\rm chan}^{\rm F} \simeq 0.019$.
Hence, the symmetry broken below the transition temperature is {\it not} simply $SU(2)_T$.
The relevant broken symmetry is actually the product $\mathscr{PT}$ where $\mathscr{T}$ is the time reversal.
Note that $\mathscr{T}$ makes
\begin{align}
\mathscr{T} \hat \Psi_{\bm{q}=\bm{0}} \mathscr{T}^{-1} &= - \hat \Psi_{\bm{q}=\bm{0}}
,\\
\mathscr{T} \hat m_{\rm chan} (\bm{q}\hspace{-1mm}=\hspace{-1mm}\bm{0}) \mathscr{T}^{-1} &= -\hat m_{\rm chan} (\bm{q}\hspace{-1mm}=\hspace{-1mm}\bm{0})
.
\end{align}
Since the product $\mathscr{PT}$ leaves $\hat m_{\rm chan}$ invariant,
${\cal H}_{\rm ext}$ in eq.~(\ref{eq_channel_field}) does not break the $\mathscr{PT}$ symmetry.   
This symmetry is broken only if $\Psi\neq 0$, which is odd under $ \mathscr{PT}$.

Away from half filling, 
generally we have $I\neq 0$ since there is no p-h symmetry.
The symmetry breaking by $\Psi$
is simply $SU(2)_T$.
If we apply ${\cal H}_{\rm ext}$ of eq.~(\ref{eq_channel_field})
away from half filling, 
we have $\Psi\neq 0$ already at higher temperature because of
the finite coupling with $m_{\rm chan}\neq 0$.
Namely the second-order transition changes into a crossover.
With weak external field, however, the channel moment in the ordered state is tiny as shown in Fig.~\ref{fig_phase}(d).
Hence, 
the crossover is rapid and should look almost like a phase transition.

\subsection{Single-particle spectrum}

\begin{figure}
\begin{center}
\includegraphics[width=85mm]{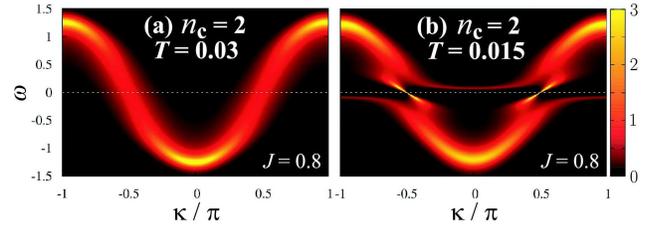}
\caption{
(Color online)
Single-particle spectra $A(\kappa, \omega)$ in (a) paramagnetic and (b) F-channel ordered states.
}
\label{fig_spectrum_half}
\end{center}
\end{figure}

Now we consider the single-particle spectrum for the F-channel order, which leads us to the effective one-body Hamiltonian at low energies.
We define the spectrum by
\begin{align}
A_{\bm{k}}(\omega) = -\frac{1}{\pi}\sum_{\alpha\sigma} \imag G_{\bm{k}\alpha\sigma} (\omega + \imu \eta)
, \label{eq_spect_def}
\end{align}
where $\eta = +0$.
The single-particle Green function is defined by
\begin{align}
G_{\bm{k}\alpha\sigma} (z) ^{-1}
= z + \mu - \varepsilon_{\bm{k}} - \Sigma _{\alpha\sigma} (z)
, 
\end{align}
where $z$ is a complex frequency and $\Sigma _{\alpha\sigma}$ is a self energy.
In the DMFT, the wave-vector dependence enters to the single-particle Green function only through $\varepsilon_{\bm{k}}$.
Hence we introduce the parameter $\kappa$ by the relation $\varepsilon_{\bm{k}} = - D \cos \kappa$, and plot the spectrum as if the system were in one dimension.

As shown in Fig.~\ref{fig_spectrum_half}(a), the spectrum in the paramagnetic state is almost the same as the non-interacting one.
The broad spectrum is consistent with the incoherent metallic state \cite{jarrell96}.
This non-Fermi liquid behavior originates from 
interaction of localized spins with overscreening degenerate channels.
In particular the Kondo singlet state cannot be formed.

On the other hand, Fig.~\ref{fig_spectrum_half}(b) shows the result in the symmetry-broken phase, where the degenerate condition of channels is no longer satisfied.
The spectrum near the Fermi level is clearly 
characterized by 
hybridized behavior of the Green function, which has the form
\begin{align}
G_{\bm{k}\alpha\sigma} (z) 
&\sim \frac{a_\alpha}{z
- a_\alpha \varepsilon_{\bm{k}}  - a_\alpha{V_\alpha}^2/z} 
, \label{eq_FL_g}\\
\Sigma_{\alpha\sigma} (z) 
&= \frac{{V_\alpha}^2}{z}
+ b_\alpha z + O\left( z^2\right)
, \label{eq_FL_sig}
\end{align}
where the renormalization factor is given by $a_\alpha = (1-b_\alpha)^{-1}$.
For $\alpha = 1$, we numerically evaluate $a_1 \simeq 0.51$ and $V_1 \simeq 0.00$ from analysis of the self energy at $T=0.005$.
This means that the conduction electrons with $\alpha = 1$ shows the Fermi-liquid behavior without hybridization.
For $\alpha = 2$, on the other hand, the values are obtained as $a_2 \simeq 0.43$, $V_2 \simeq 0.33$.
This indicates the behavior of the Kondo insulator with
effective hybridization $V_2$.
Thus, the spectrum displays the admixture of the Fermi liquid and Kondo insulator.
We can understand this behavior also from the illustration in Fig.~\ref{fig_illust}(d).
In the case away from half filling, the Kondo insulator for $\alpha = 2$ is doped into the metallic heavy Fermi liquid at sufficiently low temperatures.

The effective model that reproduces the spectrum is derived from the behaviors of the Green function (\ref{eq_FL_g}).
The low-energy behavior of Fig.~\ref{fig_spectrum_half}(b) is described by the following one-body hybridization model:
\begin{align}
{\cal H}_{\rm eff} = 
\sum_{\bm{k}\alpha\sigma} \tilde \varepsilon_{\bm{k}\alpha} c_{\bm{k}\alpha\sigma}^\dagger c_{\bm{k}\alpha\sigma}
+ {\tilde V_2} \sum_{i\sigma} (f_{i\sigma}^\dagger c_{i2\sigma} + {\rm h.c.}) 
,\label{eq_eff_hyb_half}
\end{align}
where $\tilde \varepsilon_{\bm{k}\alpha} = a_\alpha \varepsilon_{\bm{k}}$ and $\tilde V_\alpha = \sqrt{a_\alpha} V_\alpha$.
Here we regard the localized spins as ``electrons'' and introduce the fermion operator $f_{i\sigma}$.
From this effective model, it is interpreted that the localized spins acquire the itinerancy in the ordered state because of the strong renormalization by the Kondo effect.
Note that this itinerant character is absent
in the paramagnetic state; we do not see any hybridization of localized electrons as shown in Fig.~\ref{fig_spectrum_half}(a).
This is in contrast to the ordinary KL.

\section{Contrast between F-Channel and AF-Spin Orders}\label{sec_compari}

Peculiar characters of the F-channel order becomes clearer by comparing with the AF-spin order at half filling.
Roughly speaking, the F-channel and AF-spin orders resolve
the entropy $\ln\sqrt 2$ and $\ln 2$, respectively.

\subsection{Phase diagram}

\begin{figure}
\begin{center}
\includegraphics[width=75mm]{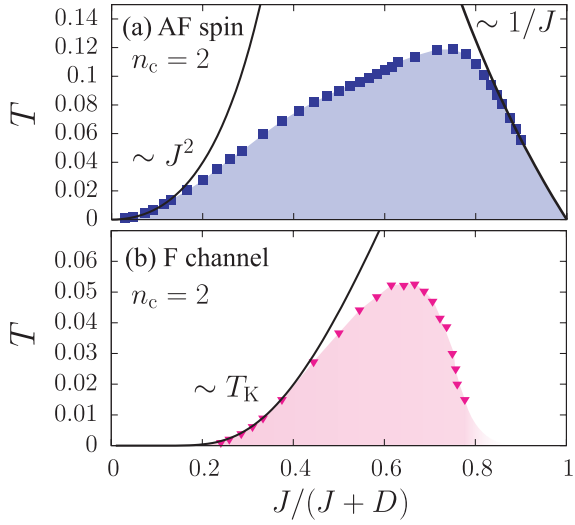}
\end{center}
\caption{(Color online)
Temperature-interaction phase diagram for the half-filled two-channel KL.
The strong coupling limit $J=\infty$ corresponds to $1$ in the horizontal axis.
}
\label{fig_phase_half}
\end{figure}

Let us first discuss the temperature-interaction phase diagram which includes the information for the origin of ordering.
Figure \ref{fig_phase_half}(a) shows the $T$ vs $J$ phase diagram for the AF-spin order at half filling.
In the weak-coupling regime, the transition temperature $T_{\rm spin}^{\rm AF}$ scales with $J^2$ as expected from the RKKY interaction.
For large couplings, on the other hand, $T_{\rm spin}^{\rm AF}$ is proportional to $1/J$, which means the strong-coupling approach account for the order.
Namely, as discussed in \S\ref{sec_phase}, the total spin $1/2$ states in $J\rightarrow\infty$ at each site are split by the effective inter-site interaction.
Thus the AF-spin order can be understood in the above two ways, that is, from either weak or strong coupling limit.

Next we consider the F-channel phase shown in Fig.~\ref{fig_phase_half}(b).
In the weak coupling region, the transition temperature scales with the Kondo temperature $T_{\rm K} \propto \exp [-1/\rho_0 (0) J]$.
Hence the weak-coupling approach fails, since $T_{\rm K}$ cannot be expanded by $J$.
For the large coupling region, on the other hand, it is likely that the transition temperature rapidly goes to zero as 
$J\rightarrow \infty$, although we cannot go to temperature region lower than $T\lesssim0.01$.
This numerical difficulty arises because the higher perturbation becomes relevant due to large $J$.
In any case, it is clear that the intermediately strong 
interaction
is essential for the F-channel ordering.

The inability to approach the F-channel ordering either from weak- or strong-coupling limit is best seen 
in the single-particle spectrum.
Namely, in the Fermi liquid channel $\alpha=1$, 
the effective Kondo coupling tends to zero, while
in the Kondo insulator channel $\alpha=2$
the coupling tends to infinity.
Thus the F-channel phase is the mixture of weak- and strong-coupling limits depending on channels, and
cannot be approached from either limit in a perturbation method.
This feature is a reminiscent of the non-trivial fixed point $J^*$ in the impurity two-channel Kondo model shown in Fig.~\ref{fig_scaling}.

\subsection{Internal energy, specific heat and entropy} 
\label{subsec_internal_ene}

\begin{figure}
\begin{center}
\includegraphics[width=85mm]{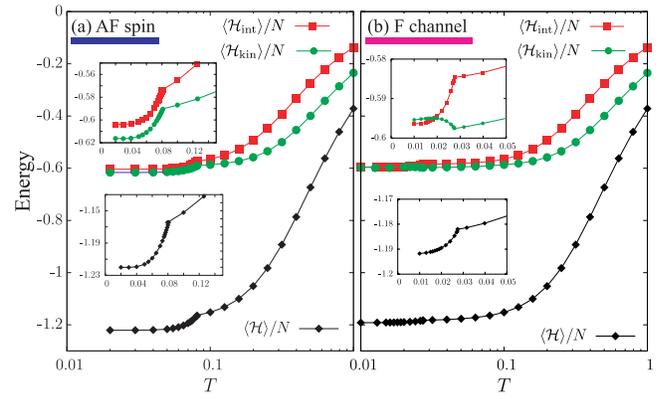}
\caption{
(Color online)
Internal energy for (a) AF-spin and (b) F-channel orders at half filling.
Kinetic energy $\langle {\cal H}_{\rm kin} \rangle$ and interaction energy $\langle {\cal H}_{\rm int} \rangle$ are also shown.
The insets show the magnified picture at low temperatures with normal scale.
}
\label{fig_internal_energy}
\end{center}
\end{figure}

We compare temperature dependence of the internal energy between the two orders at half filling.
Figure \ref{fig_internal_energy} shows the internal energy given by $\langle {\cal H}\rangle = \langle {\cal H}_{\rm kin}\rangle + \langle {\cal H}_{\rm int}\rangle$ where ${\cal H}_{\rm kin}$ and $ {\cal H}_{\rm int}$ are kinetic energy and interaction parts of the Hamiltonian.
Note that the internal energy coincides with the expectation value of the Hamiltonian because of the condition $\mu=0$ at half filling.

For the AF-spin order shown in Fig.~\ref{fig_internal_energy}(a), both the kinetic and interaction energies are lowered by the phase transition.
On the other hand, for the F-channel case in Fig.~\ref{fig_internal_energy}(b), $\langle {\cal H}_{\rm int}\rangle$ decreases below the transition temperature, while $\langle {\cal H}_{\rm kin}\rangle$ increases.
Namely the F-channel symmetry breaking is disadvantageous for conduction electrons.
Hence, we confirm that the essence of the F-channel ordering 
is in the interaction term.  Namely, the quantity (ii) in eq.~(\ref{eq_order_prms}) is not appropriate as the primary order parameter.

\begin{figure}
\begin{center}
\includegraphics[width=80mm]{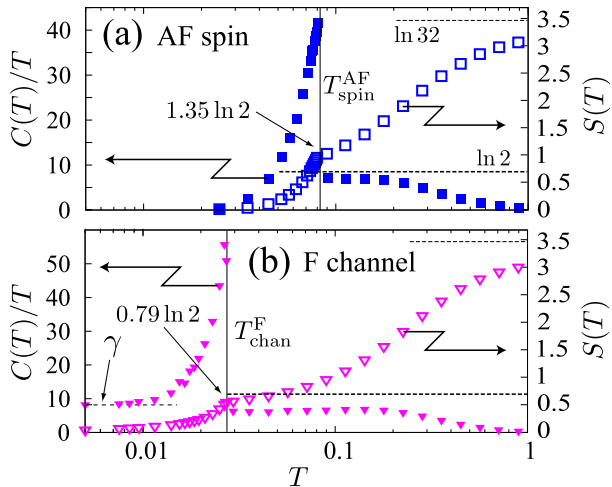}
\caption{
(Color online)
Temperature dependence of specific heat (filled symbol) and entropy (blank symbol) for (a) AF-spin and (b) F-channel orders.
The left and right scales are taken for specific heat and entropy, respectively.
The specific heat at $T=0.005$ in (b) is estimated from the renormalization factor.
}
\label{fig_specific_heat}
\end{center}
\end{figure}

Next we discuss the specific heat $C(T)$.
Figure \ref{fig_specific_heat}(a) shows the result for the AF-spin order.
The specific heat jumps at the second-order transition temperature
$T_{\rm spin}^{\rm AF}$.
Below the transition temperature, $C(T)$ goes to zero exponentially, since the energy gap is formed at the Fermi level.
For F-channel order, on the other hand, we have finite values of $C(T)/T$ at low temperatures as shown in Fig.~\ref{fig_specific_heat}(b).
The contribution comes from the Fermi-liquid channel
$\alpha = 1$  of conduction electrons.
The specific heat at low temperatures is estimated from the renormalization factor as
\begin{align}
C/T = \frac{2\pi^2\rho_0 (0)}{3a_1} \equiv \gamma
.
\end{align}
Here $a_1$ is defined in eqs.~(\ref{eq_FL_g}) and (\ref{eq_FL_sig}).
In the present case, 
we obtain $\gamma \simeq 8.2$.  As indicated in Fig.~\ref{fig_specific_heat}(b), the quasi-particle estimate is in excellent agreement with the direct numerical evaluation.

Let us consider the entropy $\Delta S$ from the ground state up to the phase transition.
By integrating the specific heat, we derive the entropy as plotted in Fig.~\ref{fig_specific_heat}.
In the high-temperature limit, the entropy tends to $S(T\rightarrow \infty) = \ln 32$ since the two-channel KL per site has 
$2\times 4^2 =32$ states with equal weights.
For the AF-spin order, we estimate as $\Delta S \simeq 1.35 \ln 2$ as shown in Fig.~\ref{fig_specific_heat}(a).
This value is interpreted as coming from the entropy $\ln 2$ of a localized spin
plus contribution from conduction electrons.

On the other hand, the F-channel order has
$\Delta S \simeq 0.79 \ln 2$ as shown in Fig.~\ref{fig_specific_heat}(b).
The value less than $\ln 2$ means that a local spin keeps correlation among themselves and with conduction electrons even in the disordered phase.
Using this result, we now derive the residual entropy in the hypothetical ground state.
The entropy change in 
the paramagnetic state from $T = 0$ to $T= T_{\rm chan}^{\rm F}$ is estimated as
$0.24\ln 2$ below $T_{\rm chan}^{\rm F}$, which is illustrated in the inset of Fig.~\ref{fig_2ch_spec}(b).
Here we have assumed that $C/T$ is nearly constant down to the ground state.
By comparing the entropy just below $T_{\rm chan}^{\rm F}$ between the hypothetical paramagnetic state and the F-channel state,  we deduce the residual entropy
$(0.79 - 0.24)\ln 2 =  0.55\ln 2$ as given in eq.~(\ref{residual}).

\section{Discussion}\label{sec_discussion}

\subsection{Relation to two-channel Kondo impurity} 
\label{subsec_majorana}

As we have seen in \S\ref{sec_para_state}, the paramagnetic state of the two-channel KL with $n_{\rm c}=2$
shows remarkable similarity to the two-channel Kondo impurity system.
This can partly be understood by the nature of the DMFT.
In each step of the DMFT iteration, we encounter an effective impurity model with self-consistently determined local density of states.  
Therefore the convergent paramagnetic solution shares the same local property with an impurity system with the corresponding density of states.
Note that the specific heat is determined by the local Green function as seen from eq.~(\ref{average-H}).
Hence it is natural that we obtain the logarithmic temperature dependence of the local susceptibility and specific heat for $n_{\rm c}\neq 2$.
What is surprising, however, is that the two-channel KL at $n_{\rm c}=2$ seems to go to the Toulouse limit\cite{emery92} for each site that contains free Majorana particles. The linear specific heat without logarithmic correction is the clearest evidence, and the residual entropy of $\ln\sqrt 2$ per site strengthens the interpretation using free Majorana particles.

We further discuss the relation between the F-channel order and Majorana particles.
According to ref.~\citen{emery92}, the relevant degrees of freedom for Majorana particles is given by the spin-channel moment
defined by eq.~(\ref{spin-channel}).
Thus the composite order parameter 
$\hat {\bm{\Psi}}_i =\bm{S}_i \cdot \hat{\bm t}_i$ 
is closely related to Majorana degrees of freedom.
In analogy with the impurity system, the residual entropy in the paramagnetic ground state is described by free Majorana fermions centered at each site.
Then it is tempting to regard
the F-channel order as a resolution of $S_0=\ln\sqrt 2$ that remains in the two-channel Kondo model.
Namely, the instability arising from the local Majorana fermions drives the system to ordering so as to release their remaining entropy toward $T=0$.

We have not yet understood why the p-h symmetry brings the system
toward the Toulouse limit
of the two-channel Kondo impurity.
The origin of nearly constant behavior of the homogeneous susceptibility $\chi (T)$ at low-temperature $T$ may also be related to the p-h symmetry.
It will be an interesting future problem to develop a perturbation theory for the two-channel KL from the limit of independent Majorana particles.

\subsection{Channel symmetry breaking described by non-Kramers doublet} \label{subsec_non-Kramers}

We discuss the relevance of our results to the non-Kramers doublet systems
where
the pseudo-spin describes a non-Kramers doublet, and the channel index $\alpha$ corresponds to the real spin.
Thus the channel-symmetry breaking in the two-channel KL implies
the breakdown of the time-reversal symmetry.

Without the p-h symmetry, the F-channel order gives rise to a finite magnetic moment corresponding to $m_{\rm chan}$ in Fig.~\ref{fig_phase}(d).
However, it is too tiny to account for the large thermodynamic anomaly in the specific heat.
The proper order parameter has been identified as the composite quantity $\hat {\bm\Psi}_i$, which has the $SU(2)_T$ symmetry but we have  considered in this paper the $z$-component.

The quantity $\hat \Psi_i^z$  itself does not give rise to a magnetic moment, although time-reversal symmetry is broken.
Hence, $\hat \Psi_i^z$ is regarded as a magnetic multipole.
This multipole moment has a radius roughly given by $v_{\rm F} / T_{\rm K}$, which is gigantic as compared to ordinary multipoles with localized character.
Namely, the order parameter is an itinerant magnetic multipole centered at each site, and the Kondo effect is responsible for the formation.
We emphasize that the composite order requires both 
itinerant and localized degrees of freedom, and is very difficult to treat
if one starts from either limit.

One may ask whether the composite order is related to
the hidden order parameter in URu$_2$Si$_2$ \cite{kuramoto09,mydosh11}.
The related system U$_x$Th$_{1-x}$Ru$_2$Si$_2$ shows the two-channel Kondo behavior \cite{amitsuka94}, and URu$_2$Si$_2$ may inherit
the property.
Then the two-channel KL can be a starting point to access URu$_2$Si$_2$.
It is likely in URu$_2$Si$_2$ that the hidden order is not homogeneous, but has a finite ordering vector.  In order to deal with this situation, we need to modify the hopping so as to include the mixing of two channels.
The higher-rank magnetic multipole in the present paper should be hard to be detected by any diffraction experiment
although the system shows the large thermodynamic and transport anomalies.
Hence the composite order parameter $\hat {\bm{\Psi}}_i$ is a candidate of the hidden order parameter.
Under the fourfold crystal symmetry,  the transverse components
$\hat \Psi_i^x$ or $\hat \Psi_i^y$
form a two-dimensional $E_u$ representation.
The doubly degenerate representation seems consistent with recently performed experiments \cite{okazaki11, takagi12}.

\subsection{Formation of narrow bands by ordering}

As seen in Fig.~\ref{fig_spectrum_half}, the narrow bands are newly formed by the phase transition into the F-channel ordered state, which is not seen in the paramagnetic phase.
Such behavior is observed also in the other ordered states such as AF-spin and AF-channel phases in the two-channel KL.

Let us compare this characteristic temperature dependence of the spectrum with that in the ordinary KL \cite{hoshino10-2}.
In the {\it ferromagnetic} KL, where the Kondo effect does not operate, the clear narrow band is never observed.
The {\it antiferromagnetic} KL, in which the Kondo effect is working, heavy bands are formed in the ordered state.
However, the clear hybridized feature of electronic spectra has already been observed in the disordered state.
This is the principal difference between single- and two-channel KLs.
This difference originates from the difficulty of forming a Kondo singlet in the two channel model.  Only by ordering, the localized spin can find a proper partner among the two.

The formation of narrow bands by ordering is recently observed in URu$_2$Si$_2$ by the single-particle spectroscopy.
In the angle-resolved photoemission spectroscopy (ARPES) \cite{yoshida10,yoshida12}, the flat band appears below the transition temperature.
On the contrary, it is not the case for the antiferromagnetic transition in a Rh-substituted sample of URu$_2$Si$_2$.
Hence the appearance of the electronic band is a characteristic of the hidden order.
A similar behavior is also observed in the scanning tunneling spectroscopy (STS) \cite{schmidt10,aynajian10,hamidian11}.
We postulate that this formation of narrow bands by ordering can be explained by the mechanism presented in this paper.

\section{Summary}\label{sec_summary}

With use of the DMFT+CTQMC approach, we have investigated the two-channel KL as the simplest model that describes periodic non-Kramers doublets coupled with conduction electrons.
In the paramagnetic state, the homogeneous susceptibility $\chi$ and specific heat $C$ at half filling show the apparent Fermi liquid  behavior.
Away from half filling, without the p-h symmetry,  both $\chi$ and $C$ show logarithmic temperature dependence.
We have derived the entropy at half filling, and found that nearly $\ln\sqrt 2$ per site remains in the paramagnetic ground state.

The channel symmetry breaking at half filling is peculiar in that the ordered state is characterized by the product of p-h symmetry $\mathscr{P}$ and the time-reversal symmetry $\mathscr{T}$.
The channel-asymmetric composite quantity between localized and conduction spins changes sign 
under this product, while the channel moment remains intact.
Thus the broken symmetry is identified as the product $\mathscr{PT}$.
The electronic spectrum shows the drastic change at the transition temperature, where the narrow bands are formed only in the ordered phase.
By comparing 
thermodynamic quantities in the F-channel and the AF-spin phases, 
we have proposed that the F-channel and AF-spin orders are characterized as resolution of entropy $\ln\sqrt 2$ and $\ln 2$, respectively.

In analogy with the two-channel Kondo impurity, the remaining entropy $S_0=\ln\sqrt  2$ in the paramagnetic state is described by the localized Majorana fermions.
The composite order is regarded as the change of these Majorana fermions into ordinary fermions by phase transition, since the ordered state is naturally described by the Fermi liquid theory.
In this paper, we have considered the Hamiltonian (\ref{eqn_2ch_KLM}) with the simple isotropic Kondo exchange coupling.
However, the form of interaction becomes more complex in general cases.
For example, the $\Gamma_5$ doublet in the tetragonal symmetry as in URu$_2$Si$_2$ has a dipole moment in addition to quadrupole moment.
Another probable situation URu$_2$Si$_2$ is that the two singlet levels are almost degenerate and form a pseudo doublet.
It is an interesting future work to study how the behaviors obtained in the two-channel KL are modified by these complications.

\acknowledgement
\section*{Acknowledgement}

We are grateful to S. Fujimoto, Y. Kato, H. Kusunose, K. Miyake and N. Nagaosa for valuable comments.
We also appreciate T. Yokoya and R. Yoshida for fruitful discussions on ARPES experiments in URu$_2$Si$_2$.
One of the authors (S.H.) acknowledges the financial support from the Japan Society for the Promotion of Science.
This work was partly supported by a Grant-in-Aid for Scientific Research on Innovative Areas ``Heavy Electrons" (No 20102008) of MEXT, Japan.
The numerical calculations were partly performed on supercomputer in the ISSP, University of Tokyo.

\appendix
\section{Numerical Evaluation of Internal Energy}

We calculate the expectation value of the Hamiltonian from the local Green function as
\begin{align}
\langle {\cal H} \rangle =
T \sum_{n} \imu \varepsilon_n G_{\rm chg} (\imu \varepsilon_n) e^{\imu \varepsilon_n \eta},
\label{average-H}
\end{align}
where $\varepsilon_n = (2n+1)\pi T$ is the fermionic Matsubara frequency.
We have defined the ``charge'' Green function by $G_{\rm chg} = \sum_{\alpha\sigma} G_{\alpha\sigma}$ where
\begin{align}
G_{\alpha \sigma} (\imu \varepsilon_n) = N^{-1} \sum_{\bm{k}} G_{\bm{k} \alpha \sigma} (\imu \varepsilon_n)
\end{align}
is the local Green function.
The internal energy $E$ is given by the relation $\langle {\cal H} \rangle = E-\mu N$.

Owing to the presence of the form factor $\imu \varepsilon_n$ in eq.~(\ref{average-H}), we have to numerically sum up a large number of Matsubara frequencies.
In order to reduce this difficulty, let us consider the expansion from high-frequency limit
\begin{align}
G_{\rm chg}(\imu \varepsilon_n) = \frac{c}{\imu \varepsilon_n} + \frac{A}{(\imu \varepsilon_n)^2} + \frac{B}{(\imu \varepsilon_n)^3} + \Delta G_{\rm chg} (\imu \varepsilon_n)
, \label{eq_ene_numerical_ver}
\end{align}
where $c=\sum_{\alpha\sigma}1=4$.
The coefficients $A$ and $B$ are real.
The 
function $\Delta G$ is introduced so that 
it behaves as $(\imu \varepsilon_n)^{-4}$ when $\varepsilon_n \rightarrow \infty$.
Then we obtain the relation
\begin{align}
\langle {\cal H} \rangle
= 
\frac{A}{2} - \frac{B}{4T} + T \sum_{n} \imu \varepsilon_n \Delta G_{\rm chg} (\imu \varepsilon_n)
. \label{eq_eval_internal_energy_numerical}
\end{align}
Note that we do not need the convergence factor in this expression.
In the numerical calculation, the summation in eq.~(\ref{eq_eval_internal_energy_numerical}) is performed in the range $|n|<L$.
We typically take $L\sim 10^4$ in the simulation.
The real coefficients $A$ and $B$ are numerically evaluated from the high-frequency behavior of the Green function as
\begin{align}
A &=  \lim_{n\rightarrow \infty} {\rm Re}\, (\imu \varepsilon_n)^2  G_{\rm chg} (\imu \varepsilon_n)
,  \label{eq_ene_coef1}\\
B &= \lim_{n\rightarrow \infty} {\rm Re}\, (\imu \varepsilon_n)^3  \left[ G_{\rm chg} (\imu \varepsilon_n) - \frac{c}{\imu \varepsilon_n} \right]
.  \label{eq_ene_coef2}
\end{align}
We have to carefully determine $A$ so that $\Delta G_{\rm chg}$ does not include the $(\imu \varepsilon_n)^{-2}$ term.
Otherwise, numerical errors can arise, since we need a convergence factor in the Matsubara-frequency summation in eq.~(\ref{eq_eval_internal_energy_numerical}).
Especially in the p-h symmetric case, $G_{\rm chg}(\imu \varepsilon_n)$ is purely imaginary, and hence the coefficient $A$ is exactly zero.


\begin{thebibliography}{99}


\bibitem{cox98} For a review, see D. L. Cox and A. Zawadowski: Adv. Phys. {\bf 47} (1998) 599.
\bibitem{amitsuka94} H. Amitsuka and T. Sakakibara: J. Phys. Soc. Jpn. {\bf 63} (1994) 736.
\bibitem{kawae05} T. Kawae, G. Li, Y. Yoshida, K. Takeda, T. Asano and T. Kitai: J. Phys. Soc. Jpn. {\bf 74} (2005) 2332.
\bibitem{onimaru10} T. Onimaru, K. T. Matsumoto, Y. F. Inoue, K. Umeo, Y. Saiga, Y. Matsushita, R. Tamura, K. Nishimoto, I. Ishii, T. Suzuki and T. Takabatake: J. Phys. Soc. Jpn. {\bf 79} (2010) 033704.
\bibitem{onimaru11} T. Onimaru, K. T. Matsumoto, Y. F. Inoue, K. Umeo, T. Sakakibara, Y. Karaki, M. Kubota and T. Takabatake: Phys. Rev. Lett. {\bf 106} (2011) 177001.
\bibitem{sakai11} A. Sakai and S. Nakatsuji: J. Phys. Soc. Jpn. {\bf 79} (2011) 063701.

\bibitem{emery92} V. J. Emery and S. Kivelson: Phys. Rev. B {\bf 46} (1992) 10812.

\bibitem{jarrell96} M. Jarrell, H. Pang, D. L. Cox and K. H. Luk: Phys. Rev. Lett. {\bf 77} (1996) 1612.
\bibitem{schauerte05} T. Schauerte D. L. Cox, R. M. Noack, P. G. J. van Dongen and C. D. Batista: Phys. Rev. Lett. {\bf 94} (2005) 147201.
\bibitem{hoshino_iche} S. Hoshino, J. Otsuki and Y. Kuramoto:
{\it Proc. Int. Conf. Heavy Electrons (ICHE2010)}
J. Phys. Soc. Jpn. {\bf 80} (2011) SA135.
\bibitem{nourafkan08} R. Nourafkan and N. Nafari: J. Phys. Condens. Matter {\bf 20} (2008) 255231.


\bibitem{hoshino11} S. Hoshino, J. Otsuki and Y. Kuramoto: Phys. Rev. Lett. {\bf 107} (2011) 247202.



\bibitem{georges96} For a review, see A. Georges, G. Kotliar, W. Krauth and M. J. Rozenberg: Rev. Mod. Phys. {\bf 68} (1996) 13.
\bibitem{gull11} For a review, see E. Gull, A. J. Millis, A. I. Lichtenstein, A. N. Rubtsov, M. Troyer and P. Werner: Rev. Mod. Phys. {\bf 83} (2011) 349.

\bibitem{hoshino10} S. Hoshino, J. Otsuki and Y. Kuramoto: Phys. Rev. B {\bf 81} (2010) 113108.


\bibitem{jarrell97} M. Jarrell, H. Pang and D. L. Cox: Phys. Rev. Lett. {\bf 78} (1997) 1996.


\bibitem{affleck92} I. Affleck, W. W. Ludwig, H. -B. Pang and D. L. Cox: Phys. Rev. B {\bf 45} (1992) 7918.
\bibitem{mitchell12} A. K. Mitchell, E. Sela and D. E. Logan: Phys. Rev. Lett. {\bf 108} (2012) 086405.
\bibitem{mitchell12-2} A. K. Mitchell and E. Sela: Phys. Rev. B {\bf 85} (2012) 235127.









\bibitem{kuramoto09} Y. Kuramoto, H. Kusunose and A. Kiss: J. Phys. Soc. Jpn. {\bf 78} (2009) 072001.
\bibitem{mydosh11} J. A. Mydosh and P. M. Oppeneer: Rev. Mod. Phys. {\bf 83} (2011) 1301.

\bibitem{okazaki11} R. Okazaki, T. Shibauchi,  H. J. Shi, Y. Haga, T. D. Matsuda, E. Yamamoto, Y. Onuki, H. Ikeda and Y. Matsuda: Science {\bf 331} (2011) 439.
\bibitem{takagi12} S. Takagi, S. Ishihara, M. Yokoyama and H. Amitsuka: J. Phys. Soc. Jpn. {\bf 81} (2012) 114710.


\bibitem{hoshino10-2} S. Hoshino, J. Otsuki and Y. Kuramoto: Phys. Rev. B {\bf 81} (2010) 113108.


\bibitem{yoshida10} R. Yoshida, Y. Nakamura, M. Fukui, Y. Haga, E. Yamamoto, Y. Onuki, M. Okawa, S. Shin, M. Hirai, Y. Muraoka and T. Yokoya: Phys. Rev. B {\bf 82} (2010) 205108.
\bibitem{yoshida12} R. Yoshida, M. Fukui, Y. Haga, E. Yamamoto, Y. Onuki, M. Okawa, W. Malaeb, S. Shin, Y. Muraoka and T. Yokoya: Phys. Rev. B {\bf 82} (2012) 241102(R).
\bibitem{schmidt10} A. R. Schmidt, M. H. Hamidian, P. Wahl, F. Meier, A. V. Balatsky, J. D. Garrett, T. J. Williams, G. M. Luke and J. C. Davis: Nature {\bf 465} (2010) 570.
\bibitem{aynajian10} P. Aynajian, E. H. da Silva Neto, C. V. Parker, Y. Huang, A. Pasupathy, J. Mydosh and A. Yazdani: Proc. Natl. Acad. Sci. U.S.A. {\bf 107} (2010) 10383.
\bibitem{hamidian11} M. H. Hamidian, A. R. Schmidt, I. A. Firmo, M. P. Allan, P. Bradley, J. D. Garrett, T. J. Williams, G. M. Luke, Y. Dubi, A. V. Balatsky and J. C. Davis: Proc. Natl. Acad. Sci. U.S.A. {\bf 108} (2011) 18233.






\end{thebibliography}
\end{document}